# One-step green synthesis of graphene nanomesh by fluid-based method

Shuaishuai Liang,[a] Min Yi,[a] Zhigang Shen,*[ab] Lei Liu,[ab] Xiaojing Zhang[a] and Shulin Ma[a]

A fluid-based method is demonstrated for preparing graphene nanomeshes (GNMs) directly from pristine graphite flakes by a one-step process. The high efficiency is attributed to the combination of fluid-assisted exfoliation and perforation of the graphene sheets. Atomic force microscopy shows that the as-produced GNMs are less than 1.5 nm thick. The total area of the pores within 1 μm² of the GNM sheet is estimated as ∼0.15 μm² and the pore density as ∼22 μm$^{-2}$, The yield of GNMs from pristine graphite powder and the power consumption for per gram GNM synthesis are evaluated as 5 wt% and 120 kW h, respectively. X-ray photoelectron spectroscopy, infrared spectroscopy, elemental analysis and Raman spectroscopy results indicate the purity of the GNMs and thus it is a green efficient method. The present work is expected to facilitate the production of GNMs in large scale.

## Introduction

Despite graphene's exceptional properties and vast potential in a wide array of applications including electronic devices,[1,2] sensors,[3–5] catalysis[6,7] and reinforced composites,[8–10] its application as a field-effect transistor (FET) working at room temperature has been hindered due to its intrinsic semi-metallic behavior with zero band gap.[11] In order to overcome this problem, researchers have tried to fabricate new graphene-based nanostructures with suitable band gaps, among which graphene nanoribbon (GNR)[12–15] and nanomesh (GNM)[11,16] are much more attractive. Although both of them can open the band gap to a level appropriate for transistor operation, GNM performs better than GNR. In fact, it has been demonstrated that the driving current or transconductance in GNM-based FET could be 100 times higher than that in individual GNR-based FET.[11] Additionally, the pore structure in GNM leads to an increase of its specific surface area and transparency, thus making GNM suitable in many applications such as catalysis, composite materials, etc. Therefore, GNM is emerging as a new fascinating nanostructure and attracting more and more attention.

However, the preparation of GNM faces challenges. So far, GNM is mainly prepared by plasma oxidation of graphene,[11,17] chemical vapor deposition,[18] or UV-assisted photodegradation of graphene oxide (GO) sheets with ZnO nanorods as the photocatalyst.[19] All the aforementioned approaches are not without drawbacks, suffering from low throughput, complexity, high cost, etc. Recently, Wang et al.[20] reported the preparation of solution-processable GNMs by refluxing reduced graphene oxide (rGO) sheets in concentrated nitric acid solution. However, the acid treatment leads to further oxidation, which introduces much more functional groups such as –C=O and –COOH to the rGO sheets.

Herein, to the best of our knowledge, we for the first time report a one-step preparation of GNMs from pristine graphite flakes by using a fluid-based method. In this method, bulk graphite particles were exfoliated into single- or few-layer graphene and were simultaneously physically punched by the cavitation-induced micro jets to form pore structures. The whole procedure causes little oxidation and is green, low cost, efficient and readily scalable.

## Experimental

The schematic of the designed device[21] used for production of GNMs is illustrated in Fig. 1. The critical part of the system is the nozzle, which is equipped with a variable cross-section flow channel for inducing cavitation and turbulence flow, as schematized in Fig. 1. Fluid-carried graphite flakes can be pressurized by an axial piston pump in the inlet, and released to ambient pressure in the outlet. In our study, natural graphite flakes were purchased from Alfa Aesar (product number 43209) and used as received. The solvent used for production of GNMs is the mixture of isopropanol and de-ionized water with a mass ratio of 1 : 1.[22] We prepared 10 L graphite dispersion by blending the natural graphite flakes in the mixed solvent at 1 mg mL$^{-1}$ and added the particle-laden fluid in the designed device to be processed for 2 h. The pressure of the inlet fluid was 30 MPa according to the pressure gauge. The resulting dispersions were centrifuged at 2000 rpm (Xiangyi L600, Changsha, China) for 1 h, and the supernatant was carefully extracted to obtain the GNM dispersion and retained for further use.

[a]Beijing Key Lab. for Powder Technology Research & Development, Beijing University of Aeronautics & Astronautics, Beijing, 100191, China. E-mail: shenzhg@buaa.edu.cn; Fax: +86-10-82338794; Tel: +86-10-82317516

[b]School of Materials Science & Engineering, Beijing University of Aeronautics & Astronautics, Beijing, 100191, China

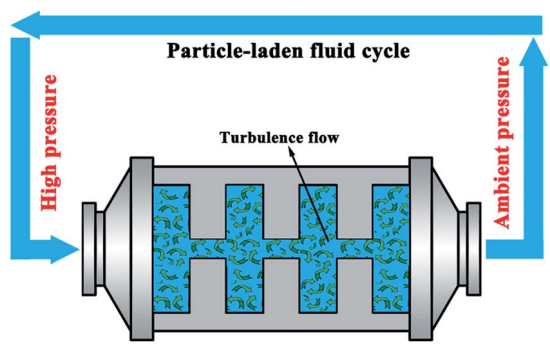

Fig. 1 Schematic of the fluid-based device used for preparing GNMs. The internal configuration of the nozzle is shown for clarity.

Optical absorbance of the GNM dispersion was measured using a Purkinje General TU1901 UV-vis spectrometer at a wavelength of 660 nm.[23] Atomic force microscope (AFM) images were collected by a Bruker MultiMode 8 scanning probe microscope in ScanAsyst Air mode. Bright-field transmission electron microscope (TEM) images were taken with a JEOL 2100 operating at 200 kV. Surface area measurement of dried graphene was performed with a Quantachrome Autosorb-IQ-MP surface and pore size analyzer using the Brunauer–Emmett–Teller (BET) method with nitrogen gas adsorption. X-ray photoelectron spectroscopy (XPS) was recorded on a Thermo Fisher Scientific ESCALAB-250 spectrometer equipped with a monochromatic Al Kα X-rays excitation source (1486.6 eV). Fourier transformer infrared (FTIR) spectrum of the GNM powder (collected from filtered films) was measured by a Nicolet Nexus 870 spectrometer using the KBr pellet technique. Elemental analysis (EA) was carried out on a vario EL cube elemental analyzer. Raman spectroscopy was captured with a Renishaw Rm2000 using a 514 nm laser. The AFM samples were prepared by dropping the GNM dispersion onto freshly cleaved mica wafer and dried in ambient temperature. Samples for TEM were made by pipetting several drops onto holey carbon mesh grid. For Raman spectroscopy, the dispersions were made into thin films by vacuum filtration through porous mixed cellulose membranes (pore size: 450 nm) and dried in ambient temperature. GNM powder for BET, XPS, FTIR, and EA study was carefully collected from the filtered films.

## Results and discussion

For the GNMs produced by fluid-based process, two critical issues should be addressed, i.e. the exfoliation state of graphene and the pore structure. With this in mind, we characterized the surface morphology of the GNMs by utilizing AFM. Fig. 2a and b show typical AFM images of the as-produced GNMs. Some pores can be clearly seen on the graphene sheets, indicating the ideal GNM structure. The height profile diagrams of Fig. 2a and b show that the thickness of the sheets was 1–1.5 nm, which can be identified as single- or bi-layer because the typical AFM-measured thickness for monolayer graphene reported by published literatures is taken as ∼1 nm.[24,25] Moreover, by conducting statistical study on 100 flakes collected from several representative AFM images, we obtained the thickness distribution of the GNMs, as shown in Fig. 2d. It can be seen that over 70% of the flakes are 1–1.5 nm thick, with others less than 1 nm. Few flakes thicker than 1.5 nm were observed. Therefore, the as-produced GNMs are proved to be highly exfoliated according to our AFM analysis. Nevertheless, the BET surface area of the

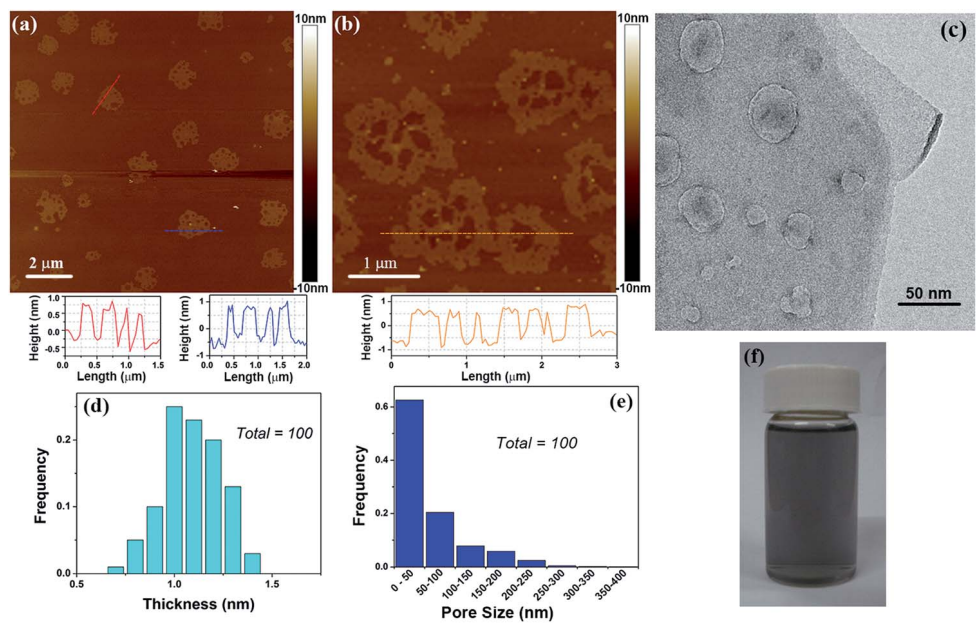

Fig. 2 (a and b) Typical AFM images of as-produced GNMs. The height profiles of corresponding lines are shown beneath the images. (c) TEM image of a GNM flake produced by the fluid-based method. (d and e) Histograms showing the distributions of thickness and pore size of the GNMs, respectively. (f) A photograph of the supernatant obtained after centrifugation, the GNM concentration was estimated as 50 μg mL$^{-1}$.

dried GNMs was measured to be 45 m$^2$ g$^{-1}$, which is significantly lower than the theoretical predictions for isolated graphene sheets (2600 m$^2$ g$^{-1}$). This indicates that the as-produced GNMs are inclined to aggregate by nature when dried down,[26] due to the absence of functional groups and wrinkle structures in its basal plane, which often exist in rGO and are critical for the separation of adjacent graphene sheets in dry state.[27] Meanwhile, the distribution of the pore sizes in GNMs was also evaluated, as presented in Fig. 2e. It can be seen that in Fig. 2e that the small pores with diameter less than 100 nm cover a large percent (~62%). Based on the statistical analysis, the total area of the pores within 1 μm$^2$ of GNM sheet can be estimated as ~0.15 μm$^2$ and the number of pores (pore density) as ~22 μm$^{-2}$. Importantly, the pore density here is comparable to that in GNMs prepared using rGO with 11 h HNO$_3$ treatment[20] and is much denser than that in the ZnO photodegradated GNMs.[19] In addition, the pore structure of the as-produced GNMs was also demonstrated by TEM analysis as shown in Fig. 2c. The TEM image clearly shows several pores in the sheet with diameters of 10–50 nm, which are in good agreement with the AFM results. The short distances of ~50 nm between neighbouring pores confirmed the pore density of the GNMs. All these results verify that the fluid-based method can produce GNMs with high pore density. Considering the important role of pore size and density in the application of GNMs,[11] our further investigation will focus on tuning the method in order to prepare GNMs with controllable pore density.

As a novel one-step method to prepare GNMs from pristine graphite, it is essential to discuss the production yield of the process. Due to that the high quality of GNMs in the supernatant (Fig. 2f) is confirmed above, we define the GNM yield as the concentration ratio of the supernatant containing GNMs (denoted as $C_S$) to the initial dispersion of pristine graphite (denoted as $C_I$), $C_S/C_I$. Here $C_S$ can be obtained according to Lambert–Beer law, given by $C_S = A/(\alpha b)$, where $A$ is the optical absorbance (measured to be 1.25), $\alpha$ is the absorption coefficient (2460 L g$^{-1}$ m$^{-1}$),[23] and $b$ is the length of the light path (1 cm), hence $C_S \sim 50$ μg mL$^{-1}$. The $C_I$ in our experiment is 1 mg mL$^{-1}$. Thus, the GNM yield is finally estimated as 5 wt%.

Another important issue we must concern is the oxidation or defect level. In the GO- or rGO-based preparation of GNM, it is of great challenge to remove the attached functional groups. For instance, according to ref. 19, GO sheets were partially reduced by the photodegradation in the presence of ZnO nanorods and further reduced by hydrazine, but the reduction was limited as indicated by the notable oxygen-containing carbonaceous bands in XPS results. In the most recent work done by Wang et al.[20] the basal-plane-related groups such as C–N, C–O in rGO sheets were reduced upon the acid treatment, but the procedure simultaneously led to a rapid increase of the –C=O and –COOH groups which are mainly located at the edges. In contrast, the fluid-based method we present here causes little oxidation as confirmed by various characterizations. Fig. 3a shows the C1s core-level XPS spectra for pristine graphite flakes and as-produced GNMs. The comparison finds no measurable change, indicating the similar chemical state for the two samples and the absence of new functional groups. The XPS survey spectra is shown in the inset of Fig. 3a, where the oxygen contents in the pristine graphite and the as-produced GNMs are labeled as 3.19% and 3.66%, respectively. Besides, the O contents measured using EA (inset of Fig. 3b) are 4.623% for GNM and 2.811% for graphite flakes, in agreement with the trend of XPS results. The slight increase of oxygen content is likely attributed to the dangling edge carbon atoms which exhibit high reactivity and are inclined to react with oxygen in the presence of water.[28,29] Furthermore, the IR spectrum of the GNMs in Fig. 3b shows no peaks associated with oxygen-containing groups, thus clearly demonstrate the absence of severe functionalization. In addition, Raman spectroscopy was used to examine structural changes and defect level of the as-produced GNMs. Typical spectra for pristine graphite flakes and GNM films prepared from dispersions by vacuum filtration are shown in Fig. 3c, where these spectra are normalized to the intensity of the G-band. It can be seen that the shape of 2D-band for GNM film is intrinsically different from that for pristine graphite flakes, indicating the exfoliation state of the GNMs.[30,31] The D-band is indicative of the presence of defects in the samples, and the defect content can be estimated by the ratio of intensity of D-band (~1350 cm$^{-1}$) to G-band (~1580 cm$^{-1}$), $I_D/I_G$. The

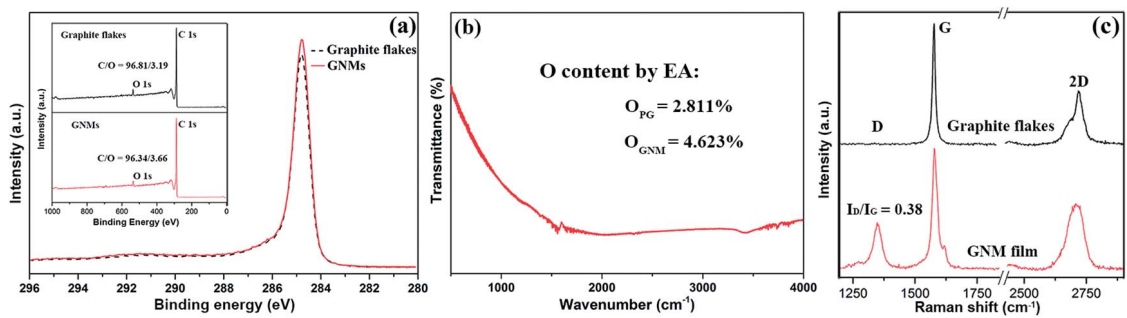

Fig. 3 (a) The C1s core-level XPS spectra for pristine graphite flakes and as-produced GNMs. Inset, XPS survey spectra for pristine graphite and GNMs. (b) FTIR spectrum of the GNM powder. Inset, EA results show the O content in pristine graphite (PG) and the GNMs. (c) Raman spectra for pristine graphite flakes and a GNM film prepared from dispersions by vacuum filtration and dried in ambient temperature. The labeled value of $I_D/I_G$ as 0.38 is averaged from 5 spots detected in the film.

defects can be divided into two main types: defects originate from edge effects and basal plane defects.[23,32] Lotya's research[33] revealed that in relatively low centrifugation rates, $I_D/I_G$ scales linearly with the ratio of mean flake edge length to area, suggesting the defects be attributed to edge effects. Accordingly, here we chose the centrifugation rate as 2000 rpm, resulted in the $I_D/I_G$ as 0.38 (averaged from 5 spots detected in the film), which falls right in the region where the defect population is dominated by edge effects.[33] Besides, the broadening of G-band which usually occur in GO was not observed in as-produced GNMs, confirming the absence of oxidative defects.[34] Therefore, the emergence of D-band here is likely due to edge effect, which is enhanced by the pore structure of GNMs.

Taking pristine graphite flakes as the precursor, the one-step formation of GNMs make the fluid-based method unique compared to other preparation methods. It is important to discuss the underlined mechanisms. Concerning this, a schematic illustration is presented in Fig. 4. In order to prepare GNMs from pristine graphite flakes, two requirements should be met, *i.e.* graphene exfoliation and perforation on graphene sheets. In the present fluid-based method, the exfoliation process can be explained by the following proposed mechanisms. Actually, there are two effects responsible for graphene exfoliation, as shown in Fig. 4. Firstly, like the widely used sonication technique, cavitation occurs in the fluid passing through the nozzle and induces micro jets and compressive shock stress waves. These shock stress waves emitted by implosion of cavitation bubbles can reach the magnitude of several MPa,[35] while the interlayer binding force of layered graphite belongs to van der Waals force and is relatively weak. Therefore, when exerting on the surface of the graphite flakes and propagating to the opposite side surface, the stress waves will reflect as normal tensile stress waves,[21] and drag graphene layers off the bulk graphite, as indicated by the green arrow in Fig. 4. The second exfoliation mechanism lies on the shear force whose direction is parallel with the graphene plane. Due to the high pressure difference across the nozzle (Fig. 1) and the sudden geometrical expansions and constrictions of its flow channel, the fluid inside the nozzle forms a turbulence flow where the velocities distribute non-homogeneously. This will result in velocity gradient between graphite flakes and the carrying fluid, thus leading to a viscous shear force which peels off thin graphene layers from the graphite flakes, as illustrated by the orange arrow in Fig. 4. The magnitude of the shear force cannot be estimated accurately because of the complexity of the velocity distribution in turbulence flow, but the exfoliation capability of the fluidic shear force has been proved by the graphene exfoliation experiment done by Chen *et al.* using a vortex fluidic device.[36] Along with the exfoliation process, perforation takes place simultaneously. As shown in Fig. 4, when the cavitation-induced micro jets exert on a tiny area in the sheet surface, they will cause punching effect and leave holes or pits on the graphene flakes. It is worthwhile to notice that the whole procedure is dominated by interactions between graphene flakes and the carrying fluid, which are intrinsically mechanical. Therefore, this method is free from chemical oxidation and functionalization, as confirmed by the previous characterizations.

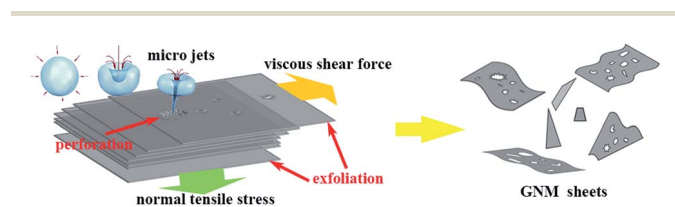

Fig. 4 A schematic illustration showing the working mechanism of the fluid-based method for preparing GNMs.

Since only energy was input during the synthesis procedure, it is necessary to evaluate the energy consumption level of the method. Typically, through 2 h preparation and the subsequent centrifugation, we could obtain 10 L GNM dispersion at a concentration of 50 μg mL$^{-1}$, *i.e.* 500 mg product. The power consumption during this process can be calculated by multiplying the nominal power of the motor (30 kW) with the operating time (2 h), 60 kW h. Thus the power consumption for per gram GNM synthesis is 120 kW h. In order to get better evaluation of the present method, we also estimated the power consumption of the widely used sonication method, where the energy output of the ultrasonic bath is 40 000 kJ for 0.7 g graphene synthesis in NMP, that is 57 000 kJ (15.8 kW h) for per gram graphene.[37] However, this is calculated using the measured power output as 23 W. When replace it with the nominal power output of the bath equipment, which is 80 W,[33,38] the final power consumption for per gram graphene synthesis should be 55.2 kW h. We acknowledge that the present method for GNM preparation consume more power than the bath sonication method. However, the method holds its own potentials for the following aspects. The concentration of the initial graphite dispersion used in our research is 1 mg mL$^{-1}$, which is more than 3 times lower than 3.3 mg mL$^{-1}$ in the sonication method. Therefore, higher yield is expected when the initial concentration is promoted. On the other hand, the present method can achieve large yield in short time process (2–4 h) rather than the prolonged sonication (*e.g.* 460 h).[37]

## Conclusions

In summary, GNMs were successfully prepared from pristine graphite flakes using the one-step fluid-based method that combines the exfoliation and perforation of graphene. The thickness and pore structure of the as-produced GNMs are clearly observed by AFM and TEM analyses, thus confirming the exfoliation and perforation ability of the process. Statistical analysis based on AFM reveals that the total area of the pores within 1 μm$^2$ of GNM sheet is ∼0.15 μm$^2$ and the number of pores ∼22. Approximately 5 wt% graphite was exfoliated and perforated during the process with a power consumption of 120 kW h for per gram GNMs synthesis. The present approach is conceptually different from those methods using GO or rGO as precursor for it introduces little oxidation as indicated by XPS, FTIR and EA results. Therefore, although some issues such as the control of pore size and density still need further investigation, the fluid-based method is expected to have great potential in providing a green, facile and scalable route in preparing GNM for a lot of applications including catalysis, batteries, supercapacitors, composite materials *etc.*

## Acknowledgements

The authors acknowledge the financial support by the Beijing Natural Science Foundation (2132025), Specialized Research Fund for the Doctoral Program of Higher Education (20131102110016), Special Funds for Co-construction Project of Beijing Municipal Commission of Education, Innovation Foundation of BUAA for Ph.D. Graduates, and Innovative Practice Foundation of BUAA for Graduates (YCSJ01201309).

## Notes and references